\documentclass[a4paper,12pt]{article}
\usepackage{amsmath,amssymb,enumerate,epsfig,bbm,theorem}
\usepackage[width=17cm,height=23cm]{geometry}

\newcommand{\dueto}[1]{\textup{\textbf{(#1) }}}
\newcommand{\tmem}[1]{{\em #1\/}}
\newcommand{\tmop}[1]{\ensuremath{\operatorname{#1}}}
\newcommand{\tmtextit}[1]{{\itshape{#1}}}
\newenvironment{enumeratenumeric}{\begin{enumerate}[1.] }{\end{enumerate}}
\newenvironment{proof}{\noindent\textbf{Proof\ }}{\hspace*{\fill}$\Box$\medskip}
\newtheorem{definition}{Definition}
{\theorembodyfont{\rmfamily}\newtheorem{example}{Example}}
\newtheorem{lemma}{Lemma}
{\theorembodyfont{\rmfamily}\newtheorem{remark}{Remark}}
\newtheorem{theorem}{Theorem}

\newcommand{\ket}[1]{\ensuremath{| #1 \rangle}}

\newcommand{\kb}[2]{\ensuremath{| #1 \rangle\!\langle #2 |}}

\begin{document}

\title{Constructing the classical limit for quantum systems on compact semisimple Lie algebras}
\author{Ingolf Sch\"afer \\
\textit{Fakult\"at f\"ur Mathematik, Ruhr-Universit\"at Bochum, } \\
\textit{D-44780 Bochum, Germany} \\
\\
Marek Ku\'s \\
\textit{Centrum Fizyki Teoretycznej PAN,}\\
\textit{ Al. Lotnik\'ow 32/42, 02-668 Warszawa,
Poland}}
\date{}
\maketitle

\begin{abstract} We give a general construction for the classical limit of a quantum
system defined in terms of generators of an arbitrary compact semisimple Lie
algebra, generalizing known results for the $\mathfrak{su}_2$ and
$\mathfrak{su}_3$ cases. The classical limit depends on the physical problem in
question and is determined by the sequence of representations by which it is
reached. Only in the simplest cases it is unique. We present explicit formulae
useful in determining the classical limit in all important cases.
\end{abstract}

\section{Introduction}

Although a natural setting for a quantum mechanical system is the
infinite-dimensional Hilbert space, there are many circumstances when the
complex projective space $\mathbb{P}^N$ suffices to describe quantum dynamics.
The most commonly known situations concern various spin systems when spatial
and spin degrees of freedom are sufficiently decoupled and we are interested
only in the evolution of the latter. In such cases, however, the notion of the
classical limit rarely makes sense, or, at least, is not immediately obvious.
There are, however, quantum systems living in finite-dimensional Hilbert spaces
(or, more correctly, their projectivisations), for which a construction of a
classical limit is sensible and conceptually simple. Let us consider a system
of $N$ atoms interacting resonantly with electromagnetic radiation. Resonance
conditions usually ensure that only a finite number $M$ of energy levels of
each atom is effectively excited by the interactions. The operators
$s_{kl}=\kb{k}{l}$ describing transitions from the level $\ket{l}$ to the level
$\ket{k}$ of a single atom, after multiplying by the imaginary unit $i$, span
the defining representation of $\mathfrak{gl}_M(\mathbb{C})$. For a system of
$N$ atoms confined to a small volume - so small that they feel the same field
amplitude - it is convenient to introduce collective operators
$S_{kl}=\sum_{\mu=1}^Ns_{kl}^\mu$ where $s_{kl}^\mu$ acts as $s_{kl}$ on the
levels of the $\mu$-th atom and as the identity on the rest. If the atoms do
not interact, the dynamics of the system can be described only in terms of
$S_{kl}$, $k,l=1,\ldots,M$ and their couplings to the electromagnetic field. It
is clear that, again up to the multiplication by $i$, $S_{kl}$, obeying the
same commutation relations as $s_{kl}$, i.e.\
$\left[S_{kl},S_{mn}\right]=\delta_{ml}S_{kn}-\delta_{kn}S_{ml}$, span a (in
general reducible) $M^N$-dimensional representation of
$\mathfrak{gl}_M(\mathbb{C})$. In a typical situation the number of atoms is
conserved and we can restrict our attention to $\mathfrak{sl}_M$. This is due
to the fact that the operator $\hat{N}=S_{11}+S_{22}+\cdots+S_{MM}$, counting
the number of atoms, is a constant of motion.

The above described example can be used to illustrate the main topic of the
present paper - a kind of classical limit for a quantum system appearing when
the number of its constituents (in our case atoms) is very large and we are
interested in such quantities like e.g.\ energy or polarization per one atom.
Such a situation in which we formally put $N\to\infty$ is common in
constructing semiclassical theory of lasers (for a particular example where
such an approach to the classical limit was applied to the so called
superradiant laser in which the active medium consisted of three-level atoms
see \cite{seeger:96}). In the limit we expect quantum operators constructed
from the generators $S_{kl}$, after scaling by the number of atoms, to be
mapped into functions defined on a classical phase space. By the classical
phase space we understand an appropriate symplectic manifold determined by the
problem in question. Moreover, to have an unambiguous connection between
quantum and classical dynamics, we demand the Dirac connection between
commutators and Poisson brackets: the classical limit of the commutator of two
operators should be proportional to the Poisson bracket of the corresponding
classical limits of the operators involved. Such a connection assures that the
classical dynamics will be obtained by applying classical limit procedure to
the quantum (Heisenberg) dynamics.

In \cite{gnutzmannkus} a detailed analysis of the problem in the case $K=3$ was
presented. The classical limit was constructed with the help of coherent states
for the group $SU_3$. (The complexification of $SU_3$, i.e.\ $SL_3(\mathbb{C})$
has, as its Lie algebra $\mathfrak{sl}_3$, generators of which appeared above
as relevant operators in the case of systems of three-level atoms). The
classical limit was obtained by taking appropriately scaled expectation values
of quantum operators in the coherent states and going to infinity with the
dimension of representation. On the manifold of the coherent states there
exists a natural symplectic structure which, in the classical limit, gives the
classical canonical structure. In the above construction we still have a
freedom to choose the way we go to infinity with the dimension of
representation. Obviously the actual way must be decided upon the actual
physical problem in question. Thus, for example in the above mentioned case of
multilevel atoms, we may consider a situation when initially the system is
prepared in the ground state, independently on the number of atoms involved.
Such a choice determines for each number of atoms a particular irreducible
representation. Increasing now the number of atoms we approach the classical
limit via a particular sequence of irreducible representations. It was shown in
\cite{haakekus} that the choice of the way to the classical limit may have
measurable consequences. In the considered case of many three level systems
there are at least two natural ways in which the classical classical limit is
achieved. The integrability properties of the system, as well as statistical
properties of spectra, distinguishing chaotic and integrable system on the
quantum level, were shown to depend on the way chosen, which makes the whole
problem interesting and worth reconsidering in the more general setting (see
also \cite{seeger:96} where the non-hamiltonian classical dynamics appropriate
for the superradiant laser is analyzed).

The present paper generalizes the above ideas to an arbitrary compact
semisimple Lie group $K$. We do not use explicitly the concept of coherent
states - a crucial element of the papers \cite{gnutzmannkus} and
\cite{haakekus}, choosing a more straightforward way of identifying classical
phase space with an appropriate orbit of the group and obtaining the classical
function with the help of the moment map. The whole construction is, however,
completely equivalent to the one using coherent states.

The crucial point consists of extending the straightforward procedure of
obtaining classical limit for generators of the Lie algebra of $K^{\mathbb{C}}$
to operators which are polynomials in the generators, as these appear in a
natural way in nontrivial quantum systems. We chose to treat such operators as
elements of the full tensor algebra $T(\mathfrak{g})$ of the Lie algebra
$\mathfrak{g}$ of $K^{\mathbb{C}}$. We show how they are represented as
differential operators, obtainable from the differential operators for the
generators, and how to perform the classical limit of them. We give formulae
useful for an arbitrary compact semisimple group which allow the explicit
calculation of classical limits for different sequences of representations.

The most important application we had in mind is the above outlined
generalization from three-level systems to ones with $M$ levels. In this case
$K=SU_M$ but, obviously, the generality of the construction gives a possibility
to analyze systems with an arbitrary compact, semisimple group of symmetries.
In a forthcoming paper we shall apply the construction to the above mentioned
problem of spectral statistics of quantum evolution operators for classically
integrable and non-integrable systems.

The two following sections of the paper give necessary definitions, conventions
and short description of properties of the momentum map for a compact group
action on a symplectic manifold (Section~\ref{momentum}) and some basic facts
from the theory of representations for compact groups
(Section~\ref{representations}. Sections~\ref{limit1} and \ref{limit2} give a
description of what is meant by a classical limit. The main results are
contained in Section~\ref{main}, where we give the formulae for calculating
classical limit for an arbitrary Lie group, generalizing results obtained in
\cite{gnutzmannkus}.  A detailed proof of an important theorem concerning the
factorization of the norm function on the orbits using methods of complex
algebraic geometry, is given in the Appendix.

\section{The momentum map}
\label{momentum}

We start this section with the definition of momentum maps which will be used
in the sequel to construct the classical limit by representation theoretic
methods.

In the simplest example of the system of free particles in $\mathbbm{R}^3$ the
momentum map assigns to each particle with a generalized momentum $p$ at a
point $q$ in configuration space the negative of its momentum, i.e.\ $-p$,
providing thus a map defined on the phase space. We shall see in the sequel
that its codomain is more interesting. Since it is a good convention, let us
call this map $\mu$.

The configuration phase space of the free particles system has translational
symmetry, so we have a natural action of the additive group of real 3-tuples,
i.e. $\mathbbm{R}^3$, by translation in configuration space. The ordinary
momentum of a particle is a point in the tangent bundle, i.e. a tangent
vector. Now we can view such a tangent vector as an infinitesimal generator of
the group $\mathbbm{R}^3$, i.e. an element of the abelian Lie algebra
$\mathbbm{R}^3$. Since generalized momenta are cotangent vectors, the momentum
map $\mu$ can be thought of as a map
\begin{equation*}
\mu:\text{phase space}\rightarrow\text{dual of Lie algebra of }\mathbbm{R}^3.
\end{equation*}
Let us discuss the abstract definition of momentum map first. We start with a
phase space $(M, \omega)$ being a smooth manifold $M$ with a symplectic
form $\omega$. A Hamiltonian vector field $\xi_H$ on $M$ is by definition a
vector field which is induced by a smooth function $H$, sometimes called the
Hamiltonian function, in such a way that the contraction of the form
$\omega$ with $\xi_H$ is the differential of $H$:
\begin{equation} i_{\xi_H} \omega = dH. \end{equation}
Now let us consider a connected Lie group $G$ of symmetries acting on $M$
\begin{equation}\label{gaction}
G\times M\rightarrow M,\quad (g,x)\mapsto g.x
\end{equation}
from the left. In order to be a group of symmetries, $G$ has to act via
canonical transformations or symplectic diffeomorphisms, i.e. for every $g\in
G$:
\begin{equation}
g^{\ast} \omega = \omega,
\end{equation}
where $g^*$ denotes the pullback by the group action (\ref{gaction}). In short
words it simply means that the symplectic structure on $M$ is invariant with
respect to the action of $G$.

What happens to the Lie algebra $\mathfrak{g}$ of $G$ here? Every $\xi \in
\mathfrak{g}$ induces a flow via $\exp (\xi t)$, which is the flow of some
vector field $r (\xi)$ on $M$. The formula for $r (\xi)$ applied to a smooth
function $f \in C^{\infty} (M)$ reads
\begin{equation}\label{eq:inducedvectorfield}
r(\xi)(f)(x)=\left. \frac{d}{d\phantom{}t}\right|_{t=0}f(\exp(-\xi t) .x),
\end{equation}
which makes $r$ a map $r : \mathfrak{g} \rightarrow \tmop{Vect} (M)$, where
$\tmop{Vect} (M)$ denotes the set of smooth vector fields on $M$. We adopted
here the sign convention which is natural for the left action, otherwise we
could choose to act on $M$ from the right and dispose the minus sign here.

Being a connected Lie group of symmetries
\begin{equation}
g^{\ast} \omega = \omega \text{ for all } g \in G
\end{equation}
is equivalent to
\begin{equation} \mathcal{L}_{r (\xi)} \omega = 0 \text{ for all }
\xi \in \mathfrak{g},
\end{equation}
where, as customary, we denote by $\mathcal{L}_X$ the Lie derivative along $X$.
By Cartan's formula we see that $\mathcal{L}_{r_{(\xi)}} \omega = d (i_{r
(\xi)} \omega) + i_{r (\xi)} d \omega$ and $d \omega = 0$, because $\omega$ as
a symplectic form is closed by definition. So, $\mathcal{L}_{r (\xi)} \omega =
d (i_{r (\xi)} \omega) = 0$ means that - at least locally - for every $\xi$
there exists  a function $\mu^{\xi}$, such that by Poincar\'e Lemma
\begin{equation}
d \mu^{\xi} = i_{r (\xi)} \omega .
\end{equation}
Let us take a closer look at the situation, pretending everything is defined
globally. For every $\xi$ we have a map
\begin{equation}
\mu^{\xi} : M \rightarrow \mathbbm{R},
\end{equation}
so we get, by choosing a $\mu^{\xi}$ for each $\xi$, a map
\begin{equation}
\mu : M \rightarrow \mathfrak{g}^{\ast}, \quad x \mapsto \mu^{(\cdot)} (x),
\end{equation}
where $\mathfrak{g}^{\ast}$ denotes the space dual to $\mathfrak{g}$.

One might wonder why $\mu^{\xi}$ should be linear in $\xi$, but since $d
\mu^{\xi} = i_{r (\xi)} \omega$ is a linear condition, it is always possible to
choose $\mu^{\xi}$ in such a way. The choice is still a non-unique one here,
because we can add any linear form $\alpha \in \mathfrak{g}^{\ast}$ to
$\mu^\xi$, i.e.
\begin{equation}
\mu^{\xi} + \alpha (\xi)
\end{equation}
is also a Hamiltonian function to the vector field $r (\xi)$.

To reduce this non-uniqueness and at the same time make a connection to the
group action, $\mu$ should be equivariant with respect to the coadjoint action
on $\mathfrak{g}^{\ast}$
\begin{equation}
g. \mu (x) = \mu (g.x),
\end{equation}
where the coadjoint action is given by
\begin{equation}
g. \alpha (\eta) = \alpha (\tmop{Ad} (g^{- 1}) .\eta) \text{ for all } \alpha \in
   \mathfrak{g}^{\ast}, \eta \in \mathfrak{g}.
\end{equation}
Thus, let us formulate the following definition
\begin{definition}{\dueto{momentum map}}
Let $(M, \omega)$ be a symplectic manifold with a Lie
group action $G \times M \rightarrow M$ by symplectic diffeomorphisms.

A momentum map is a smooth map $\mu : M \rightarrow \mathfrak{g}^{\ast}$,
which is equivariant and satisfies
\begin{equation}
d \mu^{\xi} = i_{r (\xi)} \omega \text{ for all } \xi \in \mathfrak{g}.
\end{equation}
\end{definition}

By a straightforward calculation using the flow $\exp (\xi t)$ one can
immediately prove that the equivariance property of $\mu$ is equivalent to
$\mu^{\xi}$ being a Lie algebra homomorphism, i.e.
\begin{equation}
\mu^{[\xi_1, \xi_2]} =\{\mu^{\xi_1}, \mu^{\xi_2} \}.
\end{equation}
We proceed to the next section, in which we will give an explicit formula for
the momentum map in our case.

\section{A slice of representation theory for compact groups}
\label{representations}

Since our quantum mechanical systems are given by irreducible representation
spaces of compact Lie groups, we should try to understand, at least roughly,
how the representation theory works. Details can be found in
{\cite{broecker:compactlie}}.

Let $K$ be a compact connected Lie group. Since later it will appear to be
necessary, let us assume from the beginning that $K$ is semisimple. By the
famous theorem of Peter and Weyl {\cite{peterweyl}}, we can think of $K$ as a
matrix group, more precisely $K$ is isomorphic to a subgroup of some unitary
group $U_n$.

The representation theory for $K$ relies heavily on the existence of maximal
tori, i.e.\ maximal, connected, abelian subgroups of $K$. So from now on, we
consider $K$ as a group of matrices with a fixed maximal torus $T$. We shall
denote the Lie algebra of T by $\mathfrak{t}$ and the Lie algebra of $K$ by
$\mathfrak{k}$.

Assume we have an irreducible, unitary representation{\footnote{In this article
we consider only continuous representations, so we drop the adjective
`continuous' in the following whenever it applies to a representation.}}
$\varrho$ of $K$ on some finite-dimensional Hilbert space $V$. Certainly we can
split $V$ up into irreducible subspaces with respect to the $T$-
representation:
\begin{equation}
V=\bigoplus_{\alpha}V_{\alpha}\text{ and  } \varrho =
\oplus_{\alpha} \varrho_{\alpha}
\end{equation}
The $V_\alpha$ are necessarily one-dimensional, because $T$ is abelian. So $T$
acts on these one-dimensional subspaces by multiplication with complex numbers
of norm one. These representations are called the exponentiated weights of
$\varrho$ and are given by group homomorphisms $\varrho_{\alpha} : T
\rightarrow S^1$. If we take the differential, we get a Lie algebra morphism
$d_e \varrho_{\alpha} : \mathfrak{t} \rightarrow i\mathbbm{R}$. A good
convention is to scale $d_e \varrho_{\alpha}$ a bit differently and consider
the maps $\chi_{\alpha} = \frac{1}{2 \pi i} d_e \varrho_{\alpha} : \mathfrak{t}
\rightarrow \mathbbm{R}$. These maps can be viewed as elements in the dual of
the Lie algebra $\mathfrak{t}^{\ast}$ of $\mathfrak{t}$ and we call them the
weights of the representation.

To summarize, every irreducible unitary representation $\varrho$ of K gives a
sequence of points (infinitesimal weights) in $\mathfrak{t}^{\ast}$. Since we
do not want to go through all the details, let us just say that depending on
some choice of ordering, we get a convex cone in $\mathfrak{t}^{\ast}$, usually
called the Weyl chamber, and that we can identify every irreducible, unitary
representation by the so called highest weight{\footnote{In the literature
there is normally a distinction between the weights (sometimes called real
weights) and the points in $\mathfrak{t}^{\ast}$, which are called (analytic)
integral forms. This is reasonable for a complete treatment of the matter, but
not necessary in this paper.}} in this Weyl chamber in $\mathfrak{t}^{\ast}$,
relatively to the chosen ordering. This fact is called the theorem of the
highest weight and is due to Hermann Weyl.

The set of the highest weights gives therefore a complete system of all
irreducible representations up to equivalence. Moreover, there is a natural
algebraic structure on the set of highest weights (namely it is an additive
submonoid of the additive group $\mathfrak{t}^{\ast}$ generated by a finite
number of so-called fundamental weights) - every highest weight has a unique
representation as a linear combination
\begin{equation}
\lambda = \sum \lambda_i f_i  ,
\end{equation}
with generators $f_i$ and natural numbers $\lambda_i$. (cf.
{\cite{broecker:compactlie}} chap. VI and {\cite{knapp}} Theorem 5.5)

For illustration consider the following picture of the highest weights of
$\tmop{SU}_3$ written in the coordinates $(\lambda_1, \lambda_2)$:

\begin{figure}[h]
  \begin{center}\resizebox{5cm}{!}{\epsfig{file=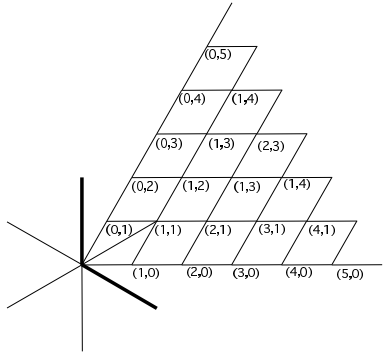}}\end{center}
  \caption{The highest weights of $\tmop{SU}_3$ written in the coordinates
  $(\lambda_1, \lambda_2)$ }
\end{figure}

Now we want to connect this to the momentum maps discussed earlier. To this
end we embed $\mathfrak{t}^{\ast}$ into $\mathfrak{k}^{\ast}$ (the dual of
$\mathfrak{k}$), with respect to the Killing form on $\mathfrak{k}$. More
formally, we identify $\mathfrak{k}$ with $\mathfrak{k}^{\ast}$ and have the
inclusion of $\mathfrak{t}^{\ast}$, if we extend every $\lambda : \mathfrak{t}
\rightarrow \mathbbm{R}$ to $\mathfrak{k}=\mathfrak{t} \oplus
\mathfrak{t}^{\perp}$ by 0 on $\mathfrak{t}^{\perp}$.

Using this identification, we can talk about $K$-orbits through $\lambda \in
\mathfrak{t}^{\ast}$. In particular we look at the coadjoint orbits through the
highest weights. It is a well-known fact that the highest weight of a
representation has multiplicity one, i.e. there is only a one-dimensional
subspace $V_{\lambda}$, such that $t.x = 2 \pi i \lambda (t) .x$ for $x \in
V_{\lambda}, t \in \mathfrak{t}$ and $\lambda \in \mathfrak{t}^{\ast}$ is the
highest weight. A vector $v_{\max} \neq 0$ in $V_{\lambda}$ is called a maximal
weight vector.

Here we encounter the basic fact:

\begin{theorem}
  Let $\varrho : K \rightarrow U (V_{\lambda})$ be an irreducible, unitary
  representation with the highest weight $\lambda$, then we have a momentum map
  $\mu : \mathbbm{P}(V_{\lambda}) \rightarrow \mathfrak{k}^{\ast}, \mu^{\xi}
  ([x]) = -2 i \frac{\langle x, d_e \varrho (\xi) .x \rangle}{\langle x, x
  \rangle}$ and $\mu ([v_{\max}]) = \lambda$. Moreover $\mu$ is a symplectic
  diffeomorphism between $K. [v_{\max}]$ with the natural symplectic form and
  $K. \lambda$ with the Kostant-Kirillov form (which is the natural symplectic
  structure on it).
\end{theorem}

\begin{proof}
  The proof that $\mu$ is a momentum map can be found in {\cite{huck:actions}}
  chapter VI.7.
   The proof that $\mu ([v_{\max}]) = \lambda$, as well as of the rest of the
   statement, is given in chapter V (Theorem 7.2) of the same reference.
\end{proof}

The natural symplectic form of the $\mathbbm{P}(V_{\lambda})$ is the one
induced by the Fubini-Study metric, i.e.\ which is the imaginary part of the
hermitian metric in $V_\lambda$ given by
\begin{equation}
  \omega = \frac{i}{2}\partial\bar{\partial} log \langle\cdot,\cdot\rangle,
\end{equation}
pushed down to $\mathbbm{P}(V_{\lambda})$.

A detailed description of Kostant-Kirillov form it is not important for the
rest of the paper. We are more interested in restating the theorem in the
following way: To every irreducible unitary representation we find a
symplectic manifold $M$ (the orbit $K. [v_{\max}]$) and a map $\tilde{\mu} :
\mathfrak{k} \rightarrow C^{\infty} (M), \xi \mapsto \mu^{\xi}$, which is a Lie
algebra homomorphism, i.e.
\begin{equation}
\tilde{\mu} ([\xi, \eta]) = \mu^{[\xi, \eta]} =\{\mu^{\xi}, \mu^{\eta} \}.
\end{equation}
This is almost a quantization procedure.

\section{The classical limit (simple case)}
\label{limit1}

We now have a connection between some skew-selfadjoint operators in the image
of $d_e \varrho_{\lambda}$ and classical functions on some symplectic manifold.
We would like, however, to have such a construction for self-adjoint operators
appearing in a natural way as observables in quantum mechanics. Clearly,
multiplying by $i$ makes a skew-selfadjoint operator selfadjoint and vice
versa. So let us consider the operators $\xi = iA$ and $\eta = iB$. We get
\begin{equation}
[\xi, \eta] = - [A, B]
\end{equation}
and
\begin{equation}
\mu^{\xi} = \mu^{iA} .
\end{equation}
We define our classical limit
\begin{equation}
\tmop{cl} : i\mathfrak{k} \rightarrow C^{\infty} (M)
\end{equation}
by
\begin{equation}
\tmop{cl} (A)(x) = \frac{1}{2}\mu^{iA}(x)=\frac{\langle  x,Ax\rangle}{\langle x,x \rangle} .
\end{equation}
Thus we get the relation
\begin{equation}
\frac{1}{2}\tmop{cl} (i [A, B]) =\{\tmop{cl} (A), \tmop{cl} (B)\},
\end{equation}
which is, modulo the constant $\hbar$, the well known Dirac quantization scheme
we wanted to have as explained in the Introduction.

\section{The classical limit (general case)}
\label{limit2}

Our classical limit has a major drawback. It is only defined for those
self-adjoint operators which are in the image of the representation $d_e
\varrho_{}$ multiplied by $i$. We would like to find a classical limit for the
system, starting with a fixed representation $\varrho$ on $V$ and some
arbitrary self-adjoint operator on $V$.

Every continuous, irreducible, unitary representation $\varrho$ of $K$ gives a
{\tmem{holomorphic}} representation of the complexification $G =
K^{\mathbbm{C}}$, which we call $\varrho$ as well. By taking the derivative we
get a representation $\varrho_{\ast} = d_e \varrho : \mathfrak{g} \rightarrow
\tmop{End} (V)$. This has a unique extension to the full tensor algebra $T
(\mathfrak{g}) = \bigoplus_{n \in \mathbbm{N}} \mathfrak{g}^{\otimes n}
 $ of $\mathfrak{g}$, by multiplying the images of generators in
$\tmop{End} (V)$.

The famous lemma of Burnside says that every operator in $\tmop{End} (V)$ is in
the image of this representation on $V$, cf. {\cite{farenick}}. This is good
and bad in some ways. Good, because we may find a way to generalize our classic
limit by algebraic methods, but bad because the procedure is not unique at all.
The non-uniqueness results from the fact that the ideal generated by the
relation $AB - BA - [A, B]$ for $A, B \in \mathfrak{g}$ is in the kernel of
every $\varrho_{_{\ast}}$. Even if we divided by this ideal, the resulting
algebra $U (\mathfrak{g})$, called the universal enveloping algebra of
$\mathfrak{g}$, would be infinite dimensional, so we would have a far greater
kernel.

To illustrate this consider for example the image of a nilpotent element
$\alpha$ in $\mathfrak{g}$. In every representation it is mapped to a
nilpotent operator. So let $A$ be the matrix representing $\alpha$ in some
suitable basis for a given representation, such that $A$ is a matrix with zero
on and below the diagonal. Then we will find some $n \in \mathbbm{N}$, such
that $A^n = 0$. This means $\alpha^n$ is in the kernel of this representation.
So, we can add $\alpha^n$ to any operator in $T (\mathfrak{g})$, without
changing the image of the representation of this operator.

To overcome this, we need a notion of hermiticity on $T (\mathfrak{g})$ and
define everything for these abstract operators independently of the chosen
representation.

\begin{definition}
  Let $\mathfrak{k}$ be the Lie algebra of a compact, simple Lie group and
  $\mathfrak{g}=\mathfrak{k} \oplus i\mathfrak{k}$ be its complexification. We
  define the formal adjoint operation $^{\dagger} : T (\mathfrak{g})
  \rightarrow T (\mathfrak{g})$ by
  \begin{enumeratenumeric}
    \item $(c \cdot 1)^{\dagger} = \bar{c} \cdot 1$

    \item $\xi^{\dagger} = - \xi$ for every $\xi \in \mathfrak{k}$

    \item $\xi^{\dagger} = \xi$ for every $\xi \in i\mathfrak{k}$

    \item $(\xi_1 \otimes \ldots \otimes \xi_n)^{\dagger} = \xi_n^{\dagger}
    \otimes \ldots \otimes \xi_1^{\dagger}$
  \end{enumeratenumeric}
  and and extend this definition to $T (\mathfrak{g})$ by linearity
  with respect to $\mathbbm{R}$.

  We call an element $\alpha \in U (\mathfrak{g})$ abstractly selfadjoint, if
  $\alpha^{\dagger} = \alpha$.
\end{definition}

\begin{remark}
  Note that the formal adjoint is not complex linear, because it contains the
  complex conjugation on the scalars in point 1.
\end{remark}

By the very definition an abstractly selfadjoint element is mapped to a
selfadjoint operator by the extended derivative $\varrho_{\ast}$ of every
irreducible unitary representation $\varrho : K \rightarrow U (V)$.

To overcome this we start with a general, abstractly selfadjoint element
$\xi_H$ in $T (\mathfrak{g})$, which we call the abstract Hamiltonian of the
system.

Let us choose some basis $\xi_1, \ldots, \xi_k$ of $\mathfrak{g}$. We can
decompose $\xi_H$ uniquely into homogeneous terms consisting of sums of
``monomials'' $\xi_{\alpha_1} \otimes \ldots \otimes \xi_{\alpha_p}$ for some
indices $\alpha_j \in \{1, \ldots, n\}$. We will extend the momentum map $\mu$
to these ``monomials'' simply by setting
\begin{equation}
\mu^{\xi_{\alpha_1} \otimes \ldots \otimes \xi_{\alpha_p}} =
   \mu^{\xi_{\alpha_1}} \cdot \ldots \cdot \mu^{\xi_{\alpha_p}},
\end{equation}
where we have already extended the definition of $\mu$ to $\mathfrak{g}$ by
complexification, that is
\begin{equation}
\mu^{\xi} ([x]) = -2 i \frac{\langle x, \varrho_{\ast} (\xi) .x
   \rangle}{\langle x, x \rangle}\quad \text{for } \xi \in \mathfrak{g}, x \in V.
\end{equation}
We extend $\mu$ by linearity to $T (\mathfrak{g})$. The resulting classical
limit $\tmop{cl}$ is physically well-behaving, because if we take only
unit vectors in $V_{\lambda}$, than
\begin{equation}
\tmop{cl}(\xi_i\xi_j)(x) =\tmop{cl}(\xi_i)(x)\tmop{cl}(\xi_j)(x)
= \langle x, \xi_i .x
\rangle \langle x, \xi_j .x \rangle,
\end{equation}
which means that the expectation value of the operator $\xi_i \xi_j$ is given by
the product of the expectation values $\xi_i$ and $\xi_j$, i.e.\ in the
classical limit $\xi_i$ and $\xi_j$ are stochastically independent operators.

This is pretty unspectacular, because it was just defined to be this way, and
we would really like so see this definition in the light of some limiting
process, sending $\hbar \rightarrow 0$. A proposal how to do this is given in
one case by direct calculation in {\cite{gnutzmannkus}}, which we will discuss
in the next chapter in general.

The reader may wonder why we do not choose the universal enveloping algebra
$U (\mathfrak{g})$ as starting point. For this consider $\xi, \xi_1, \xi_2 \in
\mathfrak{g}$, such that $0 \neq \xi = [\xi_1, \xi_2]$. In the universal
enveloping algebra we have $[\xi_1, \xi_2] = \xi_1 \xi_2 - \xi_2 \xi_1$. The
classical limit of the right hand side would be necessarily 0, because
multiplication of functions is commutative, but the left hand side is just the
classical limit of $\xi$, which is not 0. Therefore we cannot take $U
(\mathfrak{g})$ as the basic algebra of our construction.

\section{Realizing the classical limit as a mathematical limit process}
\label{main}

So far we have a generalized classical limit defined on ``monomials''
\begin{equation}
\tmop{cl} (\xi_{\alpha_1} \otimes \ldots \otimes \xi_{\alpha_p}) =
   \tmop{cl} (\xi_{\alpha_1}) \cdot \ldots \cdot \tmop{cl} (\xi_{\alpha_p})
\end{equation}
and extended by linearity.

We want to realize this limit by a limiting process, i.e.\ to have some
sequence of $C^{\infty}$-functions converging uniformly on compact sets to
the classical limit of the ``monomial''.

Our basic idea is that we should go to infinity along lines through the Weyl
chamber by which we mean the following. Assume we start in a fixed point
$\lambda$ in the Weyl chamber, which is the highest weight of some irreducible,
unitary representation $\varrho$. Let us consider the discrete line $n \cdot
\lambda$ in $\mathfrak{t}^{\ast}$ for $n \in \mathbbm{N}$. It is again a
standard fact - a corollary of the highest weight theorem - that each $n
\lambda$ is a highest weight of an irreducible, unitary representation
$\varrho_n : K \rightarrow U (V_n)$, such that for $n \neq 0$ all orbits $K. (n
\lambda)$ are diffeomorphic to each other. Moreover the isotropy subgroup is
the same for every $(n \lambda)$ and the diffeomorphisms are given by scalar
multiplications. The dimension of the irreducible representation is given by
(cf. {\cite{kirillov}})
\begin{equation}
\tmop{vol} (K. (n \lambda + \tilde{\rho})) = \dim V_n,
\end{equation}
where $\tilde{\rho}$ is a known fixed vector in the Weyl chamber and
$\tmop{vol}$ denotes the volume taken with respect to the Killing metric.

We need a little more details about the structure of $\mathfrak{g}$ now. Let us
consider the adjoint representation of $G$ on $\mathfrak{g}$. We can surely
decompose it into the one-dimensional irreducible representations of $T$. This
leads to points in $\mathfrak{t}^{\ast}$ as described in section 3. They are in
general called the weights of the representation, but since the adjoint
representation is so fundamental, its weights have a special name - they
are called the roots of the $G$. For every root $\alpha$ we find a one
dimensional subspace of $\mathfrak{g}$ called the root space to $\alpha$ and
denoted by $\mathfrak{g}_{\alpha}$.

The before mentioned ordering on $\mathfrak{t}^{\ast}$ gives a notion of
positivity on the set of roots, i.e.\ we can decompose $\mathfrak{g}$ in the
following way
\begin{equation}
\mathfrak{g}=\mathfrak{u}_- \oplus \mathfrak{t}^{\mathbbm{C}} \oplus
   \mathfrak{u}_-,
\end{equation}
where $\mathfrak{u}_+ = \sum_{\alpha \text{ positive}} \mathfrak{g}_{\alpha}
 $ and $\mathfrak{u}_- = \sum_{\alpha \text{ negative}} \mathfrak{g}_{\alpha}
$ are Lie algebras. In a good ordering of weights we can think of the matrices
in $\mathfrak{u}_- (\mathfrak{u}_+)$ \ as lower (upper) triangular matrices
with zero diagonal and $\mathfrak{t}^{\mathbbm{C}}$ as diagonal matrices. The
group $U_- = \exp (\mathfrak{u}_-)$ is biholomorphic to $\mathbbm{C}^n$,
because $\exp$ is a diffeomorphism here. Analogously, we define $U_+ = \exp
(\mathfrak{u}_+)$. For the moment it is enough to state that we have a
decomposition of the Lie algebra $\mathfrak{g}$ which gives a decomposition of
the group $G$ outside a Zariski{\footnote{Readers not familiar with the Zariski
topology may read this as closed set of lower dimension than the surrounding
space.}}-closed set, such that
\begin{equation}
G \cong \text{closure of } U_- T^{\mathbbm{C}} \phantom{} U_+ .
\end{equation}
Moreover $U_+ \subset \tmop{Stab}_G (v_{\max})$ (the stabilizer of $v_{max}$)
and $T^{\mathbbm{c}} .v_{\max} \subset \mathbbm{C}^{\ast} \cdot v_{\max}$, ie.\
$U_+$ acts trivially on $v_{\max}$ and $T^{\mathbbm{C}}$ acts on it by scalar
multiplication.

It is a general fact {\cite{huck:actions}}, that the orbit of $U_-$ through
$v_{\max}$ in $\varrho_{\lambda}$ is isomorphic to $K. [v_{\max}] \backslash A
\subset \mathbbm{P}(V_{\lambda})$, where $A$ is a Zariski-closed set in $K.
[v_{\max}]$. This leads to a chart of $K. [v_{\max}]$ given by the greatest
subgroup $U_{\max}$ of $U_-$ acting freely on $U_- .v_{\max}$. In the generic
case, that is, if the highest weight is in the interior of the Weyl chamber,
this group is $U_-$ itself. Otherwise we will have smaller unipotent groups.

The main step is to view the original classical limit $\tmop{cl}$ as
composition, $\tmop{cl}=r\circ s$, of two maps
\begin{equation}
r : i\mathfrak{k} \rightarrow \tmop{Vect} (V_{\lambda}),\, \xi \mapsto
   -\frac{1}{2}X_{\xi},\, \text{ with } (X_{\xi} f) (x) = \left. \frac{d}{dt} \right|_{t
   = 0} f (\exp (-\xi t) .x)
\end{equation}
and
\begin{equation}
s : \tmop{Vect} (V_{\lambda}) \rightarrow C^{\infty} (V_{\lambda}
   \backslash \{0\}), X \mapsto \frac{1}{N_{\lambda}} (XN_{\lambda}),
\end{equation}
where $N_{\lambda} (x) =\|x\|^2$ is the norm function squared. Note, that we
have intentionally changed our view point from $\mathbbm{P}(V_{\lambda})$ to
$V_{\lambda}$. This is a crucial step. What we will do now is not possible in
$\mathbbm{P}(V_{\lambda})$, or in physics language with normalized coherent
states, we will have to use non-normalized coherent states. The reasons for
this step will be discussed after a short calculation.

Let us thus apply the vector field $r (\xi)$ to $N_{\lambda}$ first. We have to
calculate
\begin{equation}
(X_{\xi} N_{\lambda})(x)=
\left. \frac{d}{dt}\right|_{t=0}N_{\lambda}(\exp(-\xi t).x) .
\end{equation}
Suppose that $x$ is in the $U_{\max}$-orbit through $v_{\max}$. Thus, there
exists $u \in U_{\max}$, such that
\begin{equation} x = u.v_{\max} . \end{equation}
Now we can decompose $\exp (-t \xi) u$ as $\exp (-\xi t) u = u_- (t) l (t) u_+
(t)$ for almost all{\footnote{Such a decomposition is not possible for all
elements of $G$, but on a dense, open set. This is enough for our purpose,
because we can certainly decompose for $t = 0$ and then in a small
neighborhood.}} $t$, such that $u_- (t) \in U_-$, $l (t) \in T^{\mathbbm{C}}$ and
$u_+ (t) \in U_+$.  Using the chain rule and the self-adjointness of $\xi$ we obtain
\begin{equation} (X_{\xi} N_{\lambda}) (x) =2 \langle x, \left. \frac{d}{dt}
   \right|_{t = 0} \exp (-\xi t) .x \rangle = 2\langle x, \left. \frac{d}{dt}\right|_{t = 0} u_- (t) l (t) u_+ (t) .v_{\max} \rangle . \end{equation}
Since $u_+ (t) \in U_+ \subset \tmop{Stab}_G (v_{\max})$, we have
\begin{equation} (X_{\xi} N_{\lambda}) (x) = 2\langle x, \left. \frac{d}{dt} \right|_{t = 0} u_- (t) l (t)
   .v_{\max} \rangle \end{equation}
According to the product rule we get
\begin{equation}
(X_{\xi} N_{\lambda}) (x) \left. =2\dot{l} (0) \langle x, x \rangle +
2\langle x, \frac{d}{dt} \right|_{t = 0} u_- (t) .v_{\max} \rangle,
\end{equation}
where we used the fact that the $l (t) \in T^{\mathbbm{C}}$ acts
diagonally. So, at last, we see that every vector field in the image of $r$
when applied to $N_{\lambda}$ acts like a first order differential operator,
consisting of some multiplication operator $\dot{l}$(0) and some vector field
tangential to the $U_{\max}$-orbit. The above procedure thus gives a map
\begin{equation}
\tilde{r} : i\mathfrak{k} \rightarrow \tmop{Diff} (U_{\max} .v_{\max}) .
\end{equation}
We extend $\tilde{r}$ to $T (\mathfrak{g})$ in the obvious manner, i.e.
\begin{equation}
\tilde{r} (\alpha_1 \otimes \ldots \otimes \alpha_p) = \tilde{r} (\alpha_1)
   \circ \ldots \circ \tilde{r} (\alpha_p) \text{ for } \alpha_1, \ldots,
   \alpha_p \in \mathfrak{g}
\end{equation}
and get
\begin{equation}
\tilde{r} : T (\mathfrak{g}) \rightarrow \tmop{Diff} (U_{\max} .v_{\max}) .
\end{equation}
We can now define a map $l : T (\mathfrak{g}) \rightarrow C^{\infty} (U_{\max}
.v_{\max}, \mathbbm{C}), \alpha \mapsto \frac{1}{N_{\lambda}}  \tilde{r}
(\alpha) (N_{\lambda})$ and by explicit calculation we see that $l(\alpha_j)=\tmop{cl}(\alpha_j)$.

Let us take some time to discuss more abstractly what we have done. We started
with a vector field which is tangential to the $G$-orbit through $v_{\max}$. In
the result we get a first order differential operator, which has a vector field
part tangential to the $U_{\max}$-orbit, plus some multiplication part. This is
very much like in the construction of connections on line bundles. Indeed, we
have a line bundle here. To be more precise, it is only a
$\mathbbm{C}^{\ast}$-bundle. We start with the $K$-orbit through $v_{\max}$.
Since the $G$-orbit and the $K$-orbit through $[v_{\max}]$ agree in
$\mathbbm{P}(V)$, we know that we only get points on the complex lines without
0 through the $K$-orbit.

\begin{figure}[th]
  \begin{center}\resizebox{5cm}{3cm}{\epsfig{file=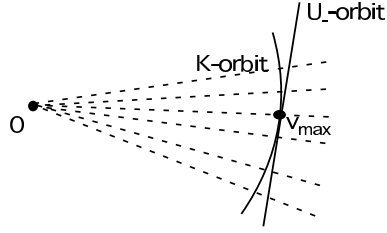}}\end{center}
  \caption{The $U_{\max}$-orbit as a section}
\end{figure}

As Figure 2 indicates, the $U_{\max}$-orbit is a
non-trivial section in this line bundle outside a closed set of lower
dimension and the multiplication part of our decomposition is just
multiplication in the fibres. We see now, why this construction is not
possible in $\mathbbm{P}(V)$. There is no such line bundle, we have
factored it out already.

Let us try to give yet another description of what happened. The $K$-orbit is
the set of normalized coherent states, while the $U_{max}$-orbit is the set of
non-normalized coherent states. If we consider some motion in the
$U_{max}$-orbit, we may project it to the $K$-orbit, but we will loose then the
information encoded in the norm of the coherent state, i.e.\ the information
which is the source of the multiplication part.

This line bundle will be of great importance later, when it comes to the proof
of the following theorem about the norm square $N_{\lambda}$:

\begin{theorem}
  \label{norm-theorem}Let $\lambda = \sum_{i = 1}^r \lambda_i f_i$ the
  decomposition of $\lambda$ into fundamental weights, then we have the
  following factorization of $N_{\lambda}$ on the $U_-$ -orbit
  \begin{equation} N_{\lambda} (x.v_{\max}) = c \cdot N_1 (x.v_{\max})^{\lambda_1} \cdot
     \ldots \cdot N_r (x.v_{\max})^{\lambda_r}, \end{equation}
  where $c$ is a real constant and $N_1, \ldots, N_r$ are the squared norms of
  the fundamental, unitary representations corresponding to the fundamental
  weights $f_1, \ldots, f_r$.
\end{theorem}

The proof of this theorem requires some methods of complex analysis and complex
algebraic geometry, which are maybe not so commonly known and can be found in
the Appendix.

\begin{theorem}\label{th3}
  Assume that $\lambda = \sum_i \lambda_i f_i$ is given, such that at least one
  $\lambda_i > p$ for a fixed positive natural number $p$.

  Let $\alpha = \xi_{\alpha_1} \otimes \ldots \otimes \xi_{\alpha_p}$ be a
  ``monomial'' element of degree $p$ in the generators $\xi_i$ of
  $\mathfrak{g}$, then $l (\alpha)$ is a polynomial of degree $p$ in the
  $\lambda_i$'s with coefficients in $C^{\infty} (U_{\max} .v_{\max},
  \mathbbm{C}$). (Recall that the $f_i$'s were a basis of the discrete cone in
  the Weyl chamber.) Moreover the homogeneous part of degree $p$ of $l
  (\alpha)$ is up to a real, multiplicative constant the product $\tmop{cl}
  (\xi_{\alpha_1}) \cdot \ldots \cdot \tmop{cl} (\xi_{\alpha_p})$. By a
  suitable choice of unitary metric this constant can be made equal to one.
\end{theorem}

\begin{proof}

  The $\tilde{r} (\xi_{\alpha_i})$ involve only first order partial
  differentiations, hence the summands in the expanded derivative of $N_{\lambda} =
  N_1^{\lambda_1} \cdot \ldots \cdot N_r^{\lambda_r}$, after dividing by
  $N_{\lambda}$, are polynomials in $\lambda$ of degree at most $p$ and there
  will be at least one summand of degree $p$. Otherwise one of the
  $\xi_{\alpha_i}$ is just the multiplication by a constant, what is not the
  case, or the partial derivatives would lower every exponent $\lambda_i$ to 0,
  what is not possible since at least one $\lambda_i$ is larger than $p$.

  Let us consider the case $p = 1$ first. In this case there will be no
  constant term, since $l (\xi_i) = \frac{1}{N_{\lambda}} r (\xi_i)
  (N_{\lambda})$ by the construction above. Now, $r (\xi_i)$ is a vector field
  and contains no multiplication term, so we have just partial derivatives
  turning $N_{\lambda}$ into a homogenous polynomial of degree one after
  dividing by $N_{\lambda}$.

  For $p \geqslant 2$ we use the product rule of differentiation. We give the
  proof here only for p=2. The rest is an exercise in systematic bookkeeping of
  indices. Let $\xi_a$ and $\xi_b$ be basis elements chosen from $(\xi_1,
  \ldots, \xi_n)$. We have $\tilde{r} (\xi_a) = \sum a_i
  \frac{\partial}{\partial z_i} + b$ for some $a_i$ and $b$ (and some
  coordinate system $\{z_i\}$ on $U_{\max} .v_{\max}$), and we can calculate $l
  (\xi_a \otimes \xi_b)$:
  \begin{eqnarray}\label{product}
    l (\xi_a \otimes \xi_b) & = & \frac{1}{N_{\lambda}} \tilde{r} (\xi_a)
    \circ \tilde{r} (\xi_b) (N_{\lambda}) = \frac{1}{N} \tilde{r} (\xi_a)
    (N_{\lambda} \cdot l (\xi_b)) \nonumber \\
    & = & \frac{1}{N_{\lambda}} (b \cdot N_{\lambda} \cdot l (\xi_b) + \sum
    a_i \frac{\partial}{\partial z_i} (N_{\lambda} \cdot l (\xi_b))\nonumber \\
    & = & l (\xi_b) \cdot \frac{1}{N_{\lambda}} \left( b \cdot N_{\lambda} +
    \sum a_i \frac{\partial}{\partial z_i} (N_{\lambda}) \right) + \sum a_i
    \frac{\partial}{\partial z_i} (l (\xi_b))\nonumber \\
    & = & l (\xi_b) \cdot l (\xi_a) + \sum a_i \frac{\partial}{\partial z_i}
    (l (\xi_b)) .
  \end{eqnarray}
  The first summand is clearly a homogenous polynomial of degree two in the
  $\lambda_i$'s, while the second lower the degree, because $l (\xi_b)$ is a
  polynomial of degree one, and taking the derivative on a polynomial lowers
  the degree by one.
\end{proof}

The formula (\ref{product}) should be compared with the explicit relations
(3.15) for the $SU_3$ case obtained in \cite{gnutzmannkus}.

With this theorem we can give an explicit construction of the classical limit
in terms of the map $l$. For the moment we have to refine our notation and use
$l_\lambda$ instead of $l$ to exhibit explicitly the dependence on the highest
weight $\lambda$. Let us now fix a given irreducible representation with the
highest weight $\lambda$, an abstract Hamiltonian $\xi_H \in T (\mathfrak{g})$
and a basis $\xi_i$ of $\mathfrak{g}$. Split $\xi_H$ into sums of monomial
elements $\alpha_i = \xi_{i_1} \otimes \ldots \otimes \xi_{i_{d (i)}}$,
$\xi_H=\sum_i\alpha_i$ and define for $n \in \mathbbm{N}$
\begin{equation}
 \tmop{cl}_n (\xi_H) = \sum_i \frac{1}{n^{d (i)}} l_{n \cdot \lambda}
   (\alpha_i)
\end{equation}
By construction and Theorem~\ref{th3}, we get
\begin{equation}
\lim_{n \rightarrow \infty} \tmop{cl}_n (\xi_H) = \sum \tmop{cl}
   (\xi_{i_1}) \cdot \ldots \cdot \tmop{cl} (\xi_{i_{d (i)}}),
\end{equation}
We denote the left hand side of the above formula by $\tmop{cl} (\xi_H)$ and
call it the classical limit of $\xi_H$. The convergence is uniform on compact
subsets of $U_- .v_{\max}$ by construction. Since the right hand side has a
smooth extension to $M$, we get our classical limit as a uniform
limit of continuous functions.

To conclude let us illustrate the construction by an example.

\begin{example}
  Consider the group $K = \tmop{SU} (3)$ of unitary $3 \times 3$ matrices. We
  have $G = K^{\mathbbm{C}} = \tmop{SL}_3 (\mathbbm{C})$, i.e.\ the unimodular
  complex $3 \times 3$ matrices. Take $$T =\Big\{\tmop{diag} (e^{i \varphi_1},
  e^{i \varphi_2}, e^{- i \varphi_1 - i \varphi_2}) | \varphi_1, \varphi_2 \in
  [0, 2 \pi [\Big\}$$ and choose $\mathfrak{u}_+$ to be the strict upper
  triangular matrices, i.e.\ with all entries below and on the diagonal equal
  to zero. $\mathfrak{u}_-$ is given then by the strict lower triangular
  matrices, while $\mathfrak{t}^{\mathbbm{C}}$ are the diagonal matrices with
  trace 0.

  The two fundamental representations of $G$ are given by the standard
  representation on $\mathbbm{C}^3$ and its dual representation. We write them
  down explicitly for $A = (a_{\tmop{ij}}) \in G$:
  \begin{eqnarray*}
    \left(\begin{array}{ccc}
      a_{11} & a_{12} & a_{13}\\
      a_{21} & a_{22} & a_{23}\\
      a_{31} & a_{32} & a_{33}
    \end{array}\right) & \mathop{\longrightarrow}\limits^{\varrho_1} &
    \left(\begin{array}{ccc}
      a_{11} & a_{12} & a_{13}\\
      a_{21} & a_{22} & a_{23}\\
      a_{31} & a_{32} & a_{33}
    \end{array}\right)
  \end{eqnarray*}
  and
  \begin{eqnarray*}
    \left(\begin{array}{ccc}
      a_{11} & a_{12} & a_{13}\\
      a_{21} & a_{22} & a_{23}\\
      a_{31} & a_{32} & a_{33}
    \end{array}\right) & \mathop{\longrightarrow}\limits^{\varrho_2} &
    \left(\begin{array}{ccc}
      \overline{a_{11}} & \overline{a_{12}} & \overline{a_{13}}\\
      \overline{a_{21}} & \overline{a_{22}} & \overline{a_{23}}\\
      \overline{a_{31}} & \overline{a_{32}} & \overline{a_{33}}
    \end{array}\right),
  \end{eqnarray*}
  where $\bar{}$ means the complex conjugate. The fact that the dual
  representation is just the complex conjugate is a consequence of the
  unitarity of $A$: its transposed inverse is equal to its complex conjugate.

  Let us consider the generator $S_{12} = \left(\begin{array}{ccc}
    0 & 1 & 0\\
    0 & 0 & 0\\
    0 & 0 & 0
  \end{array}\right)$. We would like to calculate $\tilde{r} (S_{12})$ for the
  irreducible unitary representation belonging to the highest weight
  $\lambda = (\lambda_1, \lambda_2)$ with respect to the basis $X =
  \tmop{diag} (1, - 1, 0)$ and $Y = \tmop{diag} (1, 1, - 2)$ of $\tmop{Lie}
  (T^{\mathbbm{C}}) =\mathfrak{t}^{\mathbbm{C}} .$

  In the point $x.v_{\max} = \left(\begin{array}{ccc}
    1 & 0 & 0\\
    x_{21} & 1 & 0\\
    x_{31} & x_{32} & 1
  \end{array}\right) .v_{\max}$, we have to decompose $\exp (-S_{12} t) x$:
  \begin{equation}
  \exp (-S_{12} t) x = \left(\begin{array}{ccc}
       1 & -t & 0\\
       0 & 1 & 0\\
       0 & 0 & 1
     \end{array}\right)  \left(\begin{array}{ccc}
       1 & 0 & 0\\
       x_{21} & 1 & 0\\
       x_{31} & x_{32} & 1
     \end{array}\right) =\end{equation}
  \begin{eqnarray*}
    & = & \left(\begin{array}{ccc}
      1 & 0 & 0\\
      \frac{x_{21}}{1 - t x_{21}} & 1 & 0\\
      \frac{x_{31}}{1 - t x_{21}} & x_{32} + t (x_{21} x_{32} - x_{31})
      & 1
    \end{array}\right)  \left(\begin{array}{ccc}
      1 - t x_{21} & 0 & 0\\
      0 & \frac{1}{1 - t x_{21}} & 0\\
      0 & 0 & 1
    \end{array}\right)  \left(\begin{array}{ccc}
      1 & -\frac{t}{1 - t x_{21}} & 0\\
      0 & 1 & 0\\
      0 & 0 & 1
    \end{array}\right),
  \end{eqnarray*}
  as a simple multiplication shows. Taking the derivative at $t = 0$, we get
  \begin{equation} \tilde{r} (S_{12}) = -x_{21}^2 \frac{\partial}{\partial x_{21}} - x_{21}
     x_{31} \frac{\partial}{\partial x_{31}} + (x_{31} - x_{21} x_{32})
     \frac{\partial}{\partial x_{32}} - x_{21} \lambda_1 . \end{equation}

  The general procedure applied to the considered $SU_3$ case reproduces thus
  (as promised) the results of \cite{gnutzmannkus} ).

  For the calculation of $l (S_{12})$ we need to express the $N_{\lambda}$ in
  terms of $N_1$ and $N_2$ according to Theorem~\ref{norm-theorem}:
  \begin{equation}
  N_{\lambda} = (1 + x_{21} \overline{x_{21}} + x_{31}
     \overline{x_{31}})^{\lambda_1} \cdot (1 + x_{32} \overline{x_{32}} +
     (x_{31} - x_{21} x_{32}) ( \overline{x_{31}} - \overline{x_{21}}
     \overline{x_{32}})^{\lambda_2}=N_1^{\lambda_1}N_2^{\lambda_2} .
  \end{equation}
  Applying $\tilde{r} (S_{12})$ to $N_{\lambda}$ and dividing by $N_{\lambda}$
  we get
 \begin{equation}\label{ls12}
  l (S_{12}) = \frac{x_{21}}{N_1} \lambda_1 + \frac{\overline{x_{32}}
     (x_{31} - x_{21} x_{32})}{N_2} \lambda_2
\end{equation}
  which is a polynomial in $\lambda_1$ and $\lambda_2$ with coefficients in
  $C^{\infty} (U_- .v_{\max})$, since the denominators are never zero.
\end{example}

Once more it is instructive to compare (\ref{ls12}) with Eqs.(3.18) of
\cite{gnutzmannkus} obtained in a different way.

\section{Acknowledgements}

The support by SFB/TR12 "Symmetries and Universality in Mesoscopic Systems"
program of the Deutsche Forschungsgemeischaft and Polish MEiN grant No 1P03B04226
is gratefully acknowledged.
\appendix

\section{Appendix. Proof of theorem \ref{norm-theorem}}

In this section we will give the proof of Theorem~\ref{norm-theorem}, which is
the only remaining gap to be filled in our reasoning. The essential tool is the
theorem of Borel-Weil relating a irreducible representation of a compact Lie
group to the representations on the holomorphic sections on line bundles over
flag manifolds.

Before we state the theorem we need to equip ourself with a bit of notional
convention. Let $K$ be a simple, compact Lie group with complexification $G =
K^{\mathbbm{C}}$. Let $P$ be a closed subgroup of $G$ and assume that we have
a representation $\chi : P \rightarrow \mathbbm{C}^{\ast} = \tmop{Aut}
(\mathbbm{C})$, called sometimes a multiplicative character, e.g.\ in
{\cite{huck:actions}}, but does not agree with the notion of a character of a
representation in general.

Let us briefly recall, how to get to the induced representation $\chi \uparrow G$.
The construction is the following: Consider the cartesian product of $G$ and
$\mathbbm{C}$ and define the following equivalence relation on it:
\begin{equation}
(g_1, z_1) \sim (g_2, z_2) \text{ if there exists a } p \in P \text{, such
   that } g_1 = g_2 \cdot p^{- 1} \text{ and } z_1 = \chi (p) \cdot z_2 .
\end{equation}
It is well known, that the quotient space carries the structure of a
holomorphic line bundle over $G / P$, i.e.\ if we consider the projection $\pi$
of $L  := (G \times \mathbbm{C}) / \sim$ onto the first factor, we get a well
defined map
\begin{equation}
\pi : L \rightarrow G / P,
\end{equation}
which is holomorphic with respect to the pushed down holomorphic structure on
$L$, whose fibers are one dimensional complex vector spaces. The bundle $L$ can
be trivialized by holomorphic transition functions inducing linear
transformations of the fibers. Moreover we have a natural $G - $ action on $L$
by the left multiplication on the first factor on $G \times \mathbbm{C}$ and
pushing this down to $L$, i.e.
\begin{equation}
x. [(g, z)] := [(x \cdot g, z)] .
\end{equation}
A holomorphic section of this line bundle is by definition a holomorphic map $s
: G / P \rightarrow L$, such that $\pi \circ s$=id. The space of sections of
this line bundle is a complex vector space by the pointwise addition and the
scalar multiplication, i.e.
\begin{equation}
(s_1 + s_2) (xP) := s_1 (xP) + s_2 (xP),
\end{equation}
for holomorphic sections $s_1$, $s_2$ and
\begin{equation}
(c \cdot s_1) (xP) := c \cdot s_1 (xP),
\end{equation}
for any complex number $c$.

In general this space of holomorphic sections is denoted by
$\Gamma_{\tmop{hol}} (G / P, L)$, but since we do not talk about
non-holomorphic sections we drop the subscript $_{\tmop{hol}}$ and mean that
every section should be holomorphic.

If $G / P$ is a compact complex manifold, then $\Gamma (G / P, L)$ is finite
dimensional. This is a famous statement in complex analysis, named after
Cartan and Serre. Moreover we have a representation of $G$ on $\Gamma (G / P,
L)$, which is holomorphic and given by the following linear $G$-action
\begin{equation}
(g.s) (x P) := g.s (g^{- 1} \cdot xP) .
\end{equation}
Due to the special construction of our bundle, we can identify the set of
sections in a more convenient way. Let $f : G \rightarrow \mathbbm{C}$ be a
holomorphic function, which is $P$-equivariant in the following sense
\begin{equation}
  f (gp) = \chi (p)^{- 1} f (g) . \label{aequivarianterschnitt}
\end{equation}
By a direct calculation one can show that the vector space of $P$-equivariant
functions is isomorphic to $\Gamma (G / P, L)$.

We will need only a weak version of the Borel-Weil theorem here. For a more
complete version one should look at {\cite{huck:actions}} and {\cite{akhiezer}}
for a treatment from the point of view of complex analysis. An algebraic
approach can be found in {\cite{wallachgoodman}}.

\begin{theorem}{\dueto{Borel-Weil}}

  Let $\varrho : G \rightarrow \tmop{End} (V)$ be an irreducible representation
  with highest weight $\lambda$ and $B_- := T^{\mathbbm{C}} \times U_-$ the
  Borel subgroup of the negative roots. Then $B_-$ acts by multiplication on
  $V_{\lambda}$ given by $\chi : B_- \rightarrow \mathbbm{C}^{\ast}$, where
  $\left. \left. d_e \chi \right|_{\mathfrak{t}} = 2 \pi i \lambda \right.$ and
  the representation on $\Gamma (G / B_-, L)$ is isomorphic to $\varrho$.
\end{theorem}

\begin{proof}
  For the proof cf. {\cite{akhiezer}}.
\end{proof}

For the moment let us consider the case, where the bundle $\pi : L \rightarrow
G / B_-$ is very ample, i.e. if we fix a basis $s_0, \ldots, s_N$ of $\Gamma
(G / B_-, L)$, the mapping $j : G / B_- \rightarrow \mathbbm{P}(\Gamma (G /
B_-, L))$ given by $x \mapsto [s_0 (x) : \ldots : s_N (x)]$ is a holomorphic
imbedding{\footnote{The point of the definition of very ample is that this map
is an embbeding. It is always are good map, but not necessarily an
embedding.}} of $G / B_-$ into the projective space of $\Gamma (G / B_-, L)$.
Let $\mathcal{Z}$ be the zero section of $L$. In the view of $L = G \times
\mathbbm{C}/ \sim$ the zero section is given by exactly the elements of the
form $(g, 0)$ for $g \in G$. We claim that we get an equivariant embedding of
$L \backslash \mathcal{Z}$ into $\Gamma (G / B_-, L)^{\ast}$ by the following
construction. Let $s_0^{\ast}, \ldots, s_N^{\ast}$ be the dual basis of $s_0,
\ldots, s_N$. We think of the $s_i$'s as $B_-$-equivariant functions $G
\rightarrow \mathbbm{C}$ like above and define
\begin{equation}
\varphi : L \backslash \mathcal{Z} \rightarrow \Gamma (G / B_-, L)^{\ast},
[(g, z)] \mapsto \frac{1}{z} \sum_{i = 0}^N   s_i (g) s_i^{\ast} .
\end{equation}
This is well-defined, because $z$ is not 0, otherwise $[(g, 0)] \in
\mathcal{Z}$, and independent of the choice of representative. Indeed, if we
take another representative $(gb^{- 1}, \chi (b) z)$, we get
\begin{equation}
\sum_{i = 0}^N \frac{s_i (gb^{- 1})}{\chi (b) z}   s_i^{\ast} =
\sum_{i = 0}^N   \frac{\chi (b)}{\chi (b)} \frac{s_i (g)}{z}
s_i^{\ast},
\end{equation}
because of equation (\ref{aequivarianterschnitt}).

Next, we have to show the equivariance of $\varphi$ with respect to the left
action of $G$ on $L$ and the dual representation on $\Gamma (G / B_-,
L)^{\ast}$. For this let $x^{- 1} .s_j = \sum_{i = 0}^N a_i s_i  $ for
a fixed $x \in G$ and we calculate
\begin{eqnarray*}
  x. \varphi ([g, z]) (s_j) & = & \frac{1}{z} \left( x. \sum_{i = 0}^N s_i (g)
  s_i^{\ast} \right)   (s_j)\\
  & = & \frac{1}{z} \sum_{i = 0}^N s_i (g) s_i^{\ast} (x^{- 1} .s_j)\\
  & = & \frac{1}{z} \sum_{i = 0}^N a_i s_i (g)\\
  & = & \frac{1}{z} (x^{- 1} .s_j) (g)\\
  & = & \frac{1}{z} s_j (xg)\\
  & = & \frac{1}{z} \sum_{i = 0}^N s_i (xg) s_i^{\ast}   (s_j)\\
  & = & \varphi ([xg, z]) (s_j) .
\end{eqnarray*}
Because the line bundle is very ample, $\varphi$ is an imbedding. Since
$\varphi$ is equivariant, we find a vector of maximal weight in the image
of $\varphi$ jus by taking the image of $[(e, 1)]$. This is a vector of maximal
weight, since its isotropy is exactly $B_-$, Therefore we infer, that the whole
$U_-$-orbit through every vector of maximal weight is contained in the image
of $\varphi$.

We now state the following lemma

\begin{lemma}   \label{norm-lemma}
  Let $\varphi : L \backslash \mathcal{Z} \rightarrow \Gamma (G / B_-,
  L)^{\ast}$ be the above described, equivariant imbedding. Then the unitary
  structure on $\Gamma (G / B_-, L)^{\ast}$ induces a $K$-invariant, hermitian
  bundle metric and every other $K$-invariant, hermitian bundle metric is equal
  to the former up to a constant.
\end{lemma}

\begin{proof}
  First, we state, that the $K$-action on $G / B_-$ is transitive (cf.
  {\cite{huck:actions}}), so every $K$-invariant bundle metric is the same up
  to a constant factor, so we are done once we find the induced bundle metric
  is indeed $K$-invariant.

  For this we define for $[g, z_1], [g, z_2] \in L$
  \begin{equation} h_g (z_1, z_2) = \left\{ \begin{array}{ll}
       \frac{1}{\langle \varphi ([g, z_1]), \varphi ([g, z_2]) \rangle} &
       \text{if } z_1, z_2 \neq 0\\
       0 & \text{otherwise}
     \end{array} \right. \end{equation}
  Because of the relation
  \begin{equation}
    \frac{1}{\langle \varphi ([g, z_1]), \varphi ([g, z_2]) \rangle} =
    \overline{z_1} z_2 \frac{1}{\langle \sum_{i = 0}^n f_i (g) f_i^{\ast},
    \sum_{i = 0}^n f_i (g) f_i^{\ast} \rangle} = \overline{z_1} z_2
    \frac{1}{\| \varphi ([g, 1])\|^2}, \label{buendelmetrik}
  \end{equation}
  we obtain a Hermitian inner product at every point, since $\varphi$ is
  well-defined and has values not equal to zero. We claim that $h_g$ is \ a
  smooth bundle metric. It is obvious that $h_g$ is continuous and smooth
  outside the zero section. It is a standard fact about holomorphic line
  bundles that such a bundle metric is smooth everywhere. This metric is
  $K$-invariant, because $\langle \cdot, \cdot \rangle$ is $K$-invariant and
  $\varphi$ is equivariant.

  So we get a $K$-invariant hermitian bundle metric on $L$, but since the
  $K$-action is transitive on $G / B_-$, this is unique up to a constant
  factor. This proves the statement.
\end{proof}

We need some more facts from the complex geometric representation theory. It
can happen and even will happen, that the isotropy of $[v_{\max}]$ is not
$B_+$, but a larger subgroup called then a parabolic{\footnote{A parabolic
subgroup is by definition one that contains a Borel subgroup.}} subgroup. The
line bundle $L$ over $G / B_-$ is then no longer very ample, i.e.\ $\varphi$ is
not an embedding. The solution to the problem is the following. Let us consider
an irreducible representation $\varrho$ with some highest weight $\lambda$,
such that the isotropy subgroup $P_+$ of $[v_{\max}]$ is parabolic and not
equal to $B_+$. We define the subgroup $P_-$, which corresponds to $P_+$ over
$B_+$, such that $B_- \subset P_-$ and the $P_-$ is the transposed version of
$P_+$. $P_-$ acts on $V_{\lambda}$ by multiplication with $\kappa : P_-
\rightarrow \mathbbm{C}^{\ast}$, $d_e \kappa |_{\mathfrak{t}} = 2 \pi i
\lambda$ and the representation on $\Gamma (G / P_-, G \times_{P_-}
\mathbbm{C})$ is isomorphic to $\varrho$, where we denote by $G \times_{P_-}
\mathbbm{C}$ the bundle $G \times \mathbbm{C}/ \sim$, with the equivalence
relation
\begin{equation}
(g_1, z_1) \sim (g_2, z_2) \text{ if there exists a } p \in P \text{, such
that } g_1 = g_2 \cdot p^{- 1} \text{ and } z_1 = \chi (p) \cdot z_2 .
\end{equation}
Moreover we have a fibration $f : G / B_- \rightarrow G / P_-$, such that $G
\times_{B_-} \mathbbm{C}$ is just the pullback bundle of $G \times_{P_-}
\mathbbm{C}$.

\begin{lemma}\label{tensorbuendellemma}
  Let $L_1 \rightarrow G / P'_-$ and $L_2 \rightarrow G / P''_-$ be
  homogeneous complex line bundles, that realize the representations to the
  highest weights $\lambda_1$ and $\lambda_2$. Define $P_- = P'_- \cap P''_-$.
  Then we have two fibrations $f_1 : G / P \rightarrow G / P_1$ and $f_2 : G /
  P \rightarrow G / P_2$ and the pullback bundles $f_1^{\ast} L_1$ and
  $f_2^{\ast} L_2$.

  The representation of highest weight $\lambda_1 + \lambda_2$ is then
  realized by the sections of $f^{\ast}_1 L_1 \otimes f^{\ast}_2 L_2$.
\end{lemma}

\begin{proof}
  cf. {\cite{huck:actions}}.
\end{proof}

With this preparations we can now prove Theorem~\ref{norm-theorem}.

\begin{proof}{\dueto{Theorem \ref{norm-theorem}}}
  For every fundamental representation $\varrho_{(i)}$ we have a parabolic
  subgroup $P_i$ and an equivariant embedding $\varphi_i : L_{(i), P_i}
  \backslash \mathcal{Z} \rightarrow \Gamma (G / P_i, L_{(i), P_i})$. Using
  induction and Lemma~\ref{tensorbuendellemma} we find that the representation
  with highest weight $\lambda = \sum \lambda_i f_i  $ is given by the action
  on the sections of
  \begin{equation}
  L = (f_1^{\ast} L_{(1)})^{\lambda_1} \otimes \ldots \otimes (f_r^{\ast}
  L_{(r)})^{\lambda_r},
  \end{equation}
  where the $f_i$'s are the fibrations $f_i : G / B_- \rightarrow G / P_i$.
  One should note that $B_- = P_{(1)} \cap \ldots \cap P_{(r)}$.

  It is surely enough to prove Theorem~\ref{norm-theorem} for two bundles and
  then go on by induction. Let us call these two bundles $L_a \rightarrow G /
  P_a$ and $L_b \rightarrow G / Q_b$ with highest weights $\lambda_a$ and
  $\lambda_b$ respectively. For the parabolic subgroup $P = P_a \cap P_b$ the
  line bundle $L \rightarrow G / P$, which belongs to the irreducible
  representation with highest weight $\lambda_a + \lambda_b$, is very ample.
  (cf. {\cite{huck:actions}} chap. 6.6 and 6.7).

  The base space $G / P$ fibers over $G / P_a$ and $G / P_b$ with $f_a : G / P
  \rightarrow G / P_a$ and $f_b : G / P \rightarrow G / P_b$, and the space of
  sections of the pullback bundles$f_a^{\ast} L_a$ and $f_b^{\ast} L_b$,
  realize the irreducible representations with highest weight $\lambda_a$ and
  $\lambda_b$ respectively.

  The hermitian bundle metric is also pulled back by $f^{\ast}_a$ and
  $f_b^{\ast}$ and determine again a $K$-invariant bundle metric, which is
  up to a multiplicative constant the standard one. Now, the bundle metric
  $f_a^{\ast} h_a \cdot f_b^{\ast} h_b$ is a $K$-invariant, hermitian bundle
  metric on $L = f^{\ast}_a L_a \otimes f^{\ast}_b L_b$ and agrees up to a
  skalar with the induced $K$-invariant bundle metric on $L$.

  The last step is the following: because everything is equivariant and
  homogenous, it also determines the norm up to a constant. Or to put it in
  another way, from equation (\ref{buendelmetrik}) we see that the norm is
  defined by the bundle metric up to this scalar and we apply
  Lemma~\ref{norm-lemma} here, because $L$ is very ample.

  This completes the proof of Theorem~\ref{norm-theorem}.
\end{proof}

\end{document}